\documentclass[12pt]{article}

\usepackage{amsmath,amsbsy,amscd,amssymb,graphicx,color}
\usepackage{algorithm,algorithmic}
\newcommand\NoDo{\renewcommand\algorithmicdo{}}

\newcommand\NoThen{\renewcommand\algorithmicthen{}}

\usepackage{array}
\newcolumntype{L}[1]{>{\raggedright\let\newline\\\arraybackslash\hspace{0pt}}m{#1}}
\newcolumntype{C}[1]{>{\centering\let\newline\\\arraybackslash\hspace{0pt}}m{#1}}
\newcolumntype{R}[1]{>{\raggedleft\let\newline\\\arraybackslash\hspace{0pt}}m{#1}}
\graphicspath{{figs/}}
\DeclareMathOperator*{\argmin}{arg\,min}
\newcommand{\T}{^{\text{\scriptsize{T}}}}
\begin{document}

\title{Mixed penalization in convolutive nonnegative matrix factorization for blind speech dereverberation}

\date{}

\author{
Francisco J. Ibarrola
\thanks{\footnotesize Instituto de Investigaci\'on en Se\~nales, Sistemas e Inteligencia Computacional, sinc(i), FICH-UNL/CONICET, Argentina. Ciudad Universitaria, CC 217, Ruta Nac. 168, km 472.4, (3000) Santa Fe, Argentina.
({\tt fibarrola@sinc.unl.edu.ar}).}
\and Leandro E. Di Persia
$^\ast$
\and Ruben D. Spies
\thanks{\footnotesize  Instituto de
Matem\'{a}tica Aplicada del Litoral, IMAL, CONICET-UNL, Centro Cient\'ifico Tecnol\'ogico CONICET Santa Fe, Colectora Ruta Nac. 168, km 472, Paraje ``El Pozo'', (3000), Santa Fe, Argentina and Departamento de Matem\'{a}tica, Facultad de Ingenier\'{\i}a Qu\'{\i}mica, Universidad Nacional del Litoral, Santa Fe, Argentina.} 
}

\maketitle

\begin{abstract}
When a signal is recorded in an enclosed room, it typically gets affected by reverberation. This degradation represents a problem when dealing with audio signals, particularly in the field of speech signal processing, such as automatic speech recognition. Although there are some approaches to deal with this issue that are quite satisfactory under certain conditions, constructing a method that works well in a general context still poses a significant challenge. In this article, we propose a method based on convolutive nonnegative matrix factorization that mixes two penalizers in order to impose certain characteristics over the time-frequency components of the restored signal and the reverberant components. An algorithm for implementing the method is described and tested. Comparisons of the results against those obtained with state of the art methods are presented, showing significant improvement.
\end{abstract}

\begin{center}
\normalsize
\end{center}

%===============================================
\section{Introduction}
%===============================================

In recent years, many technological developments have attracted attention towards human-machine interaction. Since the most natural and easiest way of human communication is trough speech, much research effort has been put into achieving the same natural interaction with machines. This effort has already generated many advances in a wide variety of fields such as automatic speech recognition (\cite{kim_efficient_2015}), automatic translation systems (\cite{yun_multilingual_2014}) and control of remote devices trough voice (\cite{nesselrath2016combining}), to name only a few. A significant amount of work has been recently devoted to produce robustness in speech recognition (\cite{DMR+2008}), resulting in several advances in the areas of speech enhancement (\cite{kim_efficient_2015}, \cite{martinez_denoising_2015}), multiple sources separation (\cite{DMY2009a}, \cite{di_persia_using_2016}), and particularly in dereverberation techniques (\cite{tsilfidis2010}), which constitute the topic of this work.

When  recorded in enclosed rooms, audio signals will most certainly be affected by reverberant components due to reflections of the sound waves in the walls, ceiling, floor or furniture. This can severely degrade the characteristics of the recorded signal (\cite{tashev2009}), generating difficult problems for its processing, particularly when required for certain speech applications (\cite{huang2001}). The goal of any dereverberation technique is to remove or to attenuate the reverberant components in order to obtain a cleaner signal. The dereverberation problem is called ``blind'' when the available data consists only of the reverberant signal itself, and this is the problem we shall deal with in this work.

Depending on the problem, our observation might consist of a single or multi-channel signal. That is, we might have a signal recorded by one or more microphones. For the latter case, quite a few methods exist that work relatively well (\cite{delcroix2014}, \cite{wisdom2014}).

For the single-channel case, we may distinguish between supervised and unsupervised approaches. The first kind refers to those that begin with a training stage that serves to learn some characteristics of the reververation conditions, while the second kind alludes to those methods that can be implemented directly over the reverberant signal. Some supervised methods (\cite{moshirynia2014}, \cite{xiao2014}, \cite{nathwani2015}) appear to perform somewhat better than unsupervised ones, but they pose the disadvantage of needing learning data corresponding to the specific room conditions, microphone and source locations, and a previous process that might take a significant amount of time.

In the context of unsupervised blind dereverberation, although some recently proposed methods (\cite{wisdom2014}, \cite{kameoka2009}) work reasonably well, there is still much room for improvement. Our work is based on a convolutive non-negative matrix factorization (NMF) reverberation model, as proposed by Kameoka \emph{et al} (\cite{kameoka2009}), along with a Bayesian approach for building a generalized functional that mixes two types of penalizers over the elements of the representation model. Mixed penalization approaches have been recently used and successfully applied by several authors in many areas, mainly in signal and image processing applications (\cite{ibarrola2017l2bva}, \cite{lazzaro2015}, \cite{ibarrola2017inp}, \cite{peterson2017}, \cite{mazzieri2015}). These techniques have shown to produce good results in terms of enhancing certain desirable characteristics on the solutions while precluding unwanted ones.

%---------------------------------------------------------------------------------------------
\subsection{A Reverberation Model}
%---------------------------------------------------------------------------------------------

Let $s, x:\mathbb{R}\rightarrow\mathbb{R}$, with support in $[0, \infty)$, be the functions associated to the clean and reverberant signals, respectively. As it is customary, we shall assume that the reverberation process is well represented by a Linear Time-Invariant (LTI) system. Thus, the reverberation model can be written as
\begin{equation} \label{eq:cont-model}
x(t) = (h\ast s)(t),
\end{equation}
where $h:\mathbb{R}\rightarrow\mathbb{R}$ is the room impulse response (RIR) signal, and ``$\ast$'' denotes convolution. This LTI hypothesis implies we are assuming the source and microphone positions to be static, and the energy of the signal to be low enough for the effect of the non-linear components to be relatively insignificant.

When dealing with sound signals (particularly speech signals), it is often convenient to work with the associated spectrograms rather than the signals themselves. Thus, we make use of the short time Fourier transform (STFT), defined as
\begin{equation}\nonumber%\label{eq:STFTcont}
\mathbf{x}_k(t) \doteq \int_{-\infty}^{\infty}x(u)w(u-t)e^{-2\pi i u k}du,\;\;t,k\in\mathbb{R},
\end{equation}
where $w:\mathbb{R}\rightarrow\mathbb{R}^+_0$ is a compactly supported, even function such that $\|w\|_1 = 1$. This function is called \emph{window}.

In practice, we work with discretized versions of the signals involved ($x[\cdot],h[\cdot],s[\cdot],$ and $w[\cdot]$). With this in mind, we shall define the discrete STFT as
\begin{equation}\nonumber
\mathbf{x}_k[n] \doteq \sum_{m =-\infty}^{\infty}x[m]w[m-n]e^{-2\pi i m k},\;\;n,k\in\mathbb{N}.
\end{equation}

Denoting the STFTs of $s$ and $h$ by $ \mathbf{s}_k[n]$ and $\mathbf{h}_k[n]$, respectively, a discretized approximation of the STFT model associated to (\ref{eq:cont-model}) is given by
\begin{equation}
\mathbf{x}_k[n] \approx \tilde{\mathbf{x}}_k[n] \doteq \sum_{\tau = 0}^{N_h-1} \mathbf{s}_k[n-\tau] \mathbf{h}_k[\tau],
\end{equation}
where $n=1,\ldots, N,$ is a discretized time variable that corresponds to window location, $k=1,\ldots, K,$ denotes the frequency subband and $N_h$ is a parameter of the model associated to the expected maximum duration of the reverberation phenomenon. The model is built as in \cite{avargel2007system}, being the approximation due to the use of badn-to-band filters only. Later on, the values of $n$ will be chosen in such a way that the union of the windows' supports contain the support of the observed signal, and the values of $k$ in such a way that they cover the whole frequency spectrum, up to half the sampling frequency.

Now, let us write $\mathbf{h}_k[\tau] = |\mathbf{h}_k[\tau]|e^{j\phi_k[\tau]}$. It is well known (\cite{yegnanarayana1998}) that the phase angles $\phi_k[\tau]$ are highly sensitive with respect to mild variations on the reverberation conditions. To overcome the problems derived from this, we shall proceed (see \cite{kameoka2009}) treating the $K\times N_h$ variables $\phi_k[\tau]$ as \emph{i.i.d.} random variables  with uniform distribution in $[-\pi,\pi)$. Denoting the complex conjugate by ``$^*$'' and the Kronecker delta by $\delta_{ij}$, the expected value of $|\tilde{\mathbf{x}}_k[t]|^2$ is given by
\begin{align}
E|\tilde{\mathbf{x}}_k[n]|^2 &= E \sum_{\tau,\tau'} \mathbf{s}_k[n-\tau] \mathbf{s}_k^*[n-\tau'] \mathbf{h}_k[\tau] \mathbf{h}^*_k[\tau'] \nonumber\\
&= E \sum_{\tau,\tau'} \mathbf{s}_k[n-\tau] \mathbf{s}_k^*[n-\tau'] \,|\mathbf{h}_k[\tau]|\,e^{j\phi_k[\tau]}\, |\mathbf{h}_k[\tau']|\,e^{-j\phi_k[\tau']} \nonumber\\
&= \sum_{\tau,\tau'} \mathbf{s}_k[n-\tau] \mathbf{s}_k^*[n-\tau']\, |\mathbf{h}_k[\tau]| \,|\mathbf{h}_k[\tau']|\,Ee^{j(\phi_k[\tau]-\phi_k[\tau'])} \nonumber\\
&= \sum_{\tau,\tau'} \mathbf{s}_k[n-\tau] \mathbf{s}_k^*[n-\tau']\, |\mathbf{h}_k[\tau]|\, |\mathbf{h}_k[\tau']|\, \delta_{\tau \tau'} \nonumber\\
&= \sum_{\tau} |\mathbf{s}_k[n-\tau]|^2 \,|\mathbf{h}_k[\tau]|^2. \nonumber
\end{align}

Note that the $[-\pi,\pi)$ interval choice for $\phi_k[\tau]$ is arbitrary, since this result holds for any $2\pi-$length interval. Finally, let us define $S_k[n] \doteq |\mathbf{s}_k[n]|^2$, $H_k[n] \doteq |\mathbf{h}_k[n]|^2$ and $X_k[n] \doteq E|\tilde{\mathbf{x}}_k[n]|^2$. Then, our model reads
\begin{equation} \label{eq:mod-rep}
X_k[n] = \sum_{\tau} S_k[n-\tau] H_k[\tau],
\end{equation}
and the square magnitude of the observed spectrogram components can be written as
\begin{equation} \label{eq:mod-obs}
Y_k[n] = X_k[n] +\epsilon_k[n],
\end{equation}
where $\epsilon_k[n]$ denotes the representation error. As shown in \cite{kameoka2009}, this model is equivalent to a convolutive NMF (\cite{smaragdis2004}) with diagonal basis. In the next section, we derive a cost function in order to find an appropriate convolutive representation that allows us to isolate the components $S_k[n]$.

%===============================================
\section{A Bayesian approach}
%===============================================

In the following, we will use a Bayesian approach to derive a cost function which we will then minimize in order to obtain our regularized solution. Let us begin by assuming, for every $k$,  $\epsilon_k[n], S_k[n], H_k[n]$ are independent random variables, also independent with respect to $k$. Also, let us denote by  $S,Y, X\in \mathbb{R}^{K\times N}$ and $H \in \mathbb{R}^{K\times N_h}$ the non-negative matrices whose $(k,n)$-th elements are $S_k[n],Y_k[n], X_k[n]$ and $H_k[n]$, respectively.

As it is customary  (\cite{kameoka2009}), for the representation error, we assume $\epsilon_k[n] \sim \mathcal{N}(0,\sigma^2)$, where $\sigma>0$ is an unknown parameter, and the variables are non-correlated with respect to $n$. Hence, it follows from (\ref{eq:mod-obs}) that the conditional distribution of $Y$ given $S$ and $H$ (i.e. the likelihood distribution) is given by
\begin{equation}
\pi_{like}(Y|S,H) = \prod_{k=1}^K \prod_{n=1}^N  \frac{1}{\sqrt{2\pi}\sigma} \exp \left(-\frac{(Y_k[n] - X_k[n])^2}{\sigma^2}\right). \nonumber
\end{equation}

\begin{figure} [h]
	\centering
\includegraphics[width=0.49\textwidth]{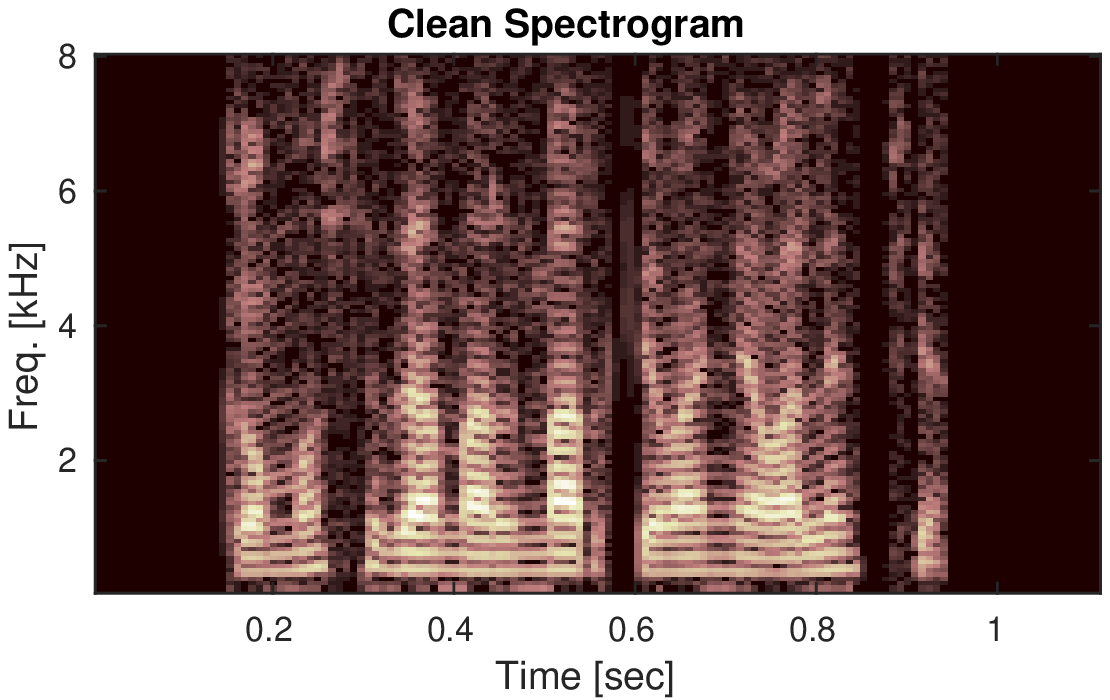}
\includegraphics[width=0.49\textwidth]{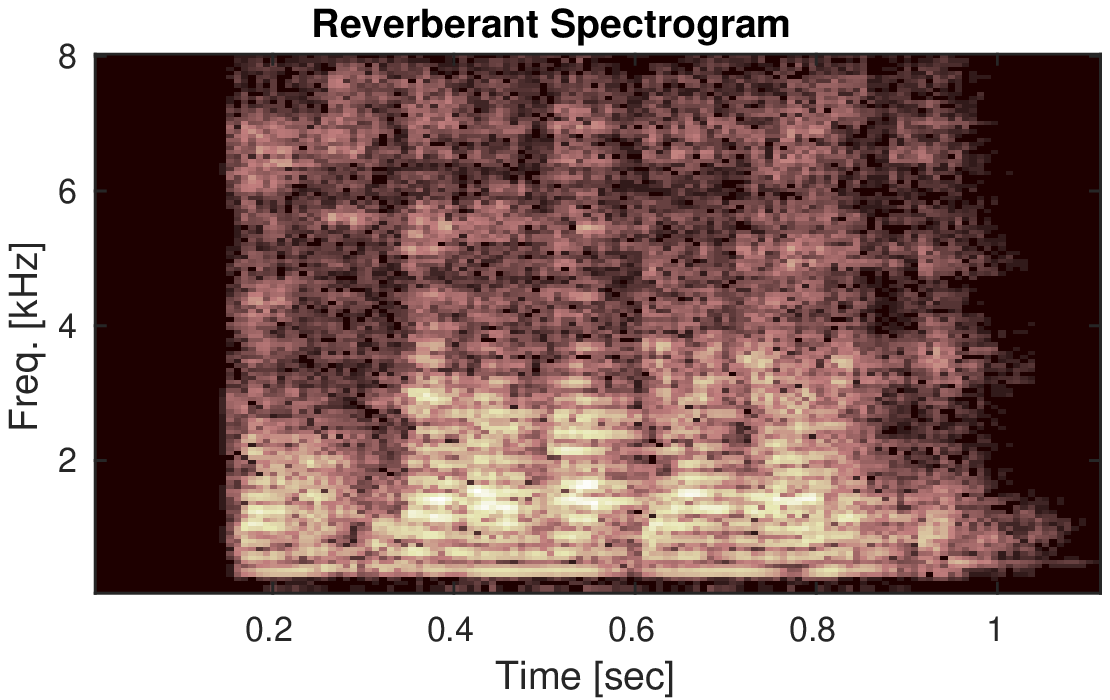}
\caption{Spectrograms for a clean speech signal (left) and the corresponding reverberant speech signal (right).}
\label{fi:clean-rev}
\end{figure}

Let us now turn our attention to $S$. Figure \ref{fi:clean-rev} depicts the $\log$-spectrograms for a clean signal and its reverberant version. As it can be observed, while the spectrogram of the clean signal is somewhat sparse, the one corresponding to the reverberant signal presents a smoother or more diffuse structure. The presence of  discontinuities in the spectrogram of the clean signal can be favored by assuming $S$ follows a generalized Gaussian distribution (\cite{bouman1993}). Namely,
\begin{equation}
\pi_{prior}(S) =\prod_{k=1}^K \prod_{n=1}^N  \frac{1}{2\Gamma(1+1/p)b_k} \exp \left(-\frac{|S_k[n]|^p}{b_k^p}\right), \nonumber
\end{equation}
where $p\in(0,2)$ is a prescribed parameter and $b_k>0$ is unknown.

\begin{figure} [h]
	\centering
\includegraphics[width=0.32\textwidth]{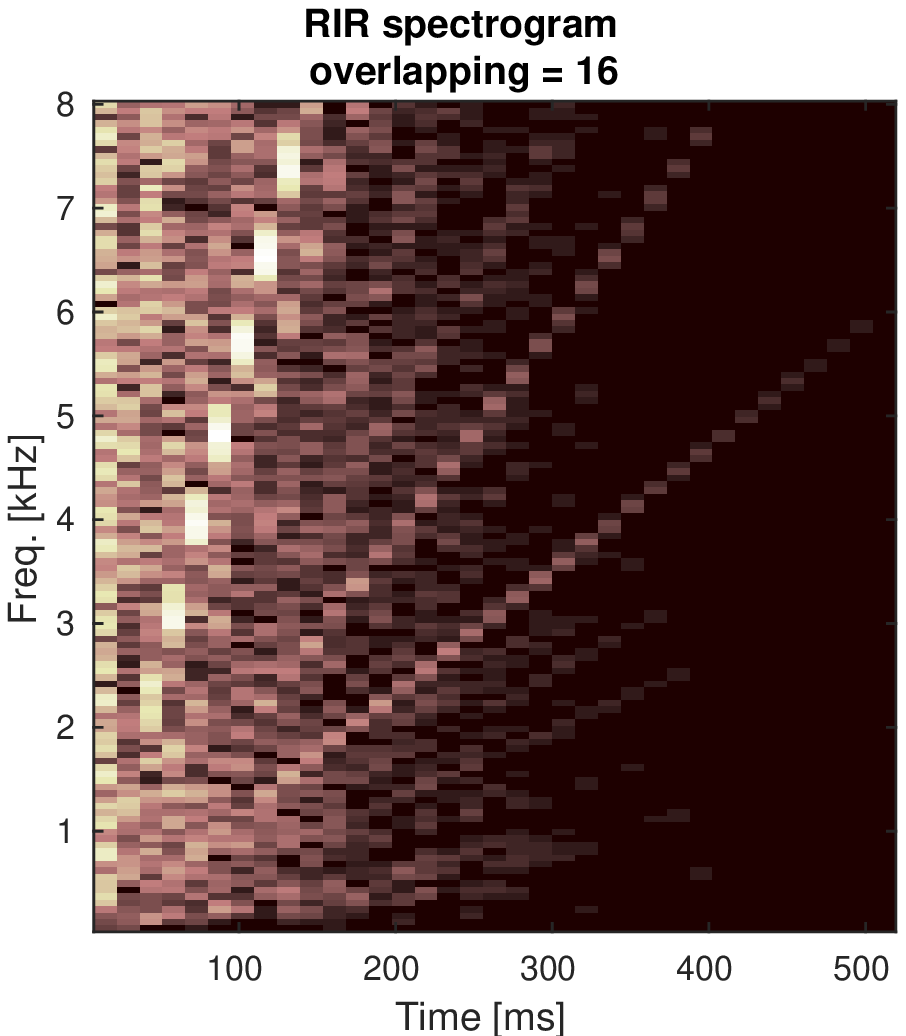}
\includegraphics[width=0.32\textwidth]{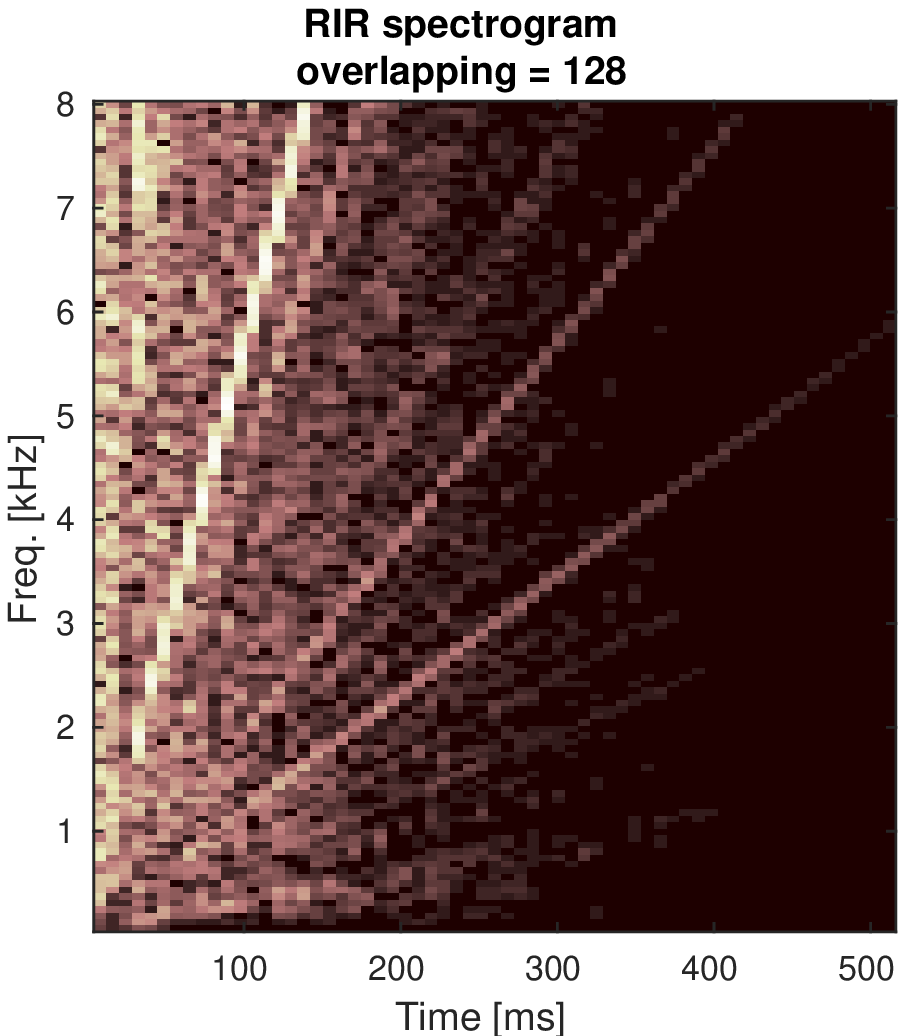}
\includegraphics[width=0.32\textwidth]{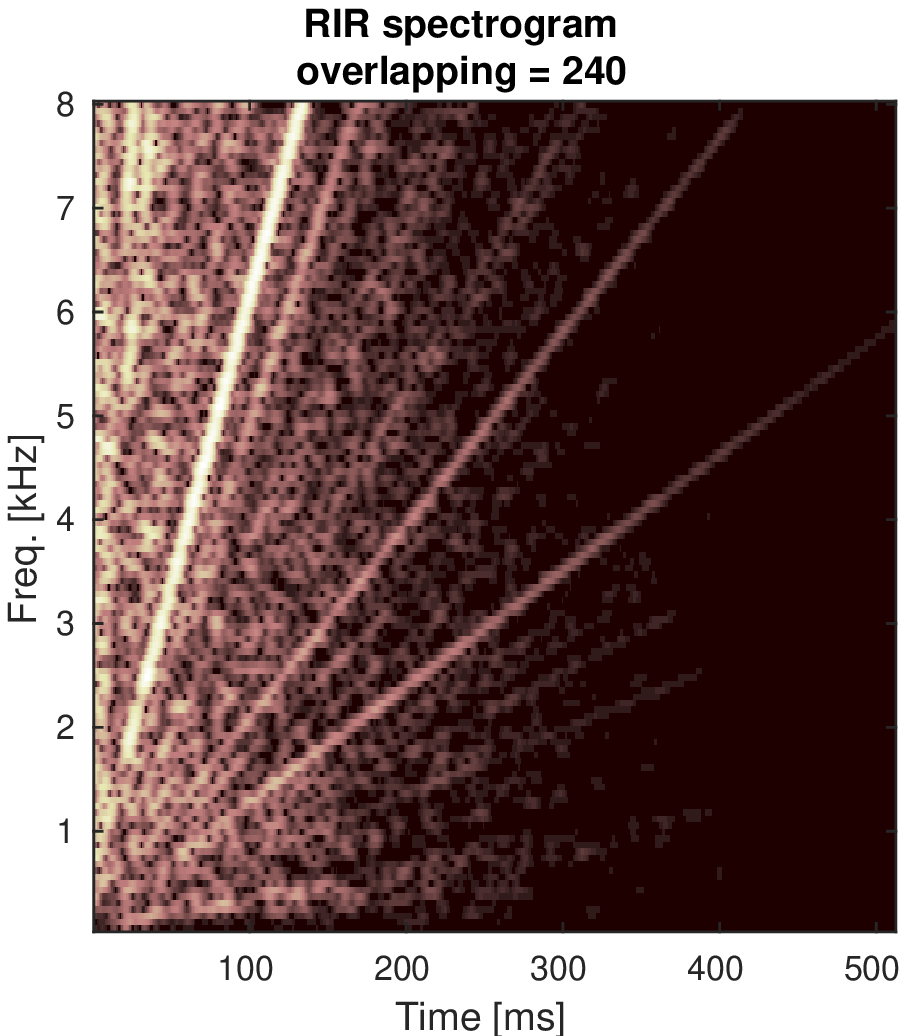}
\caption{Log-spectrograms for a RIR signal with window length 256 and different overlappings.}
\label{fi:rirspecs}
\end{figure}

\begin{figure} [h]
	\centering
\includegraphics[width=\textwidth]{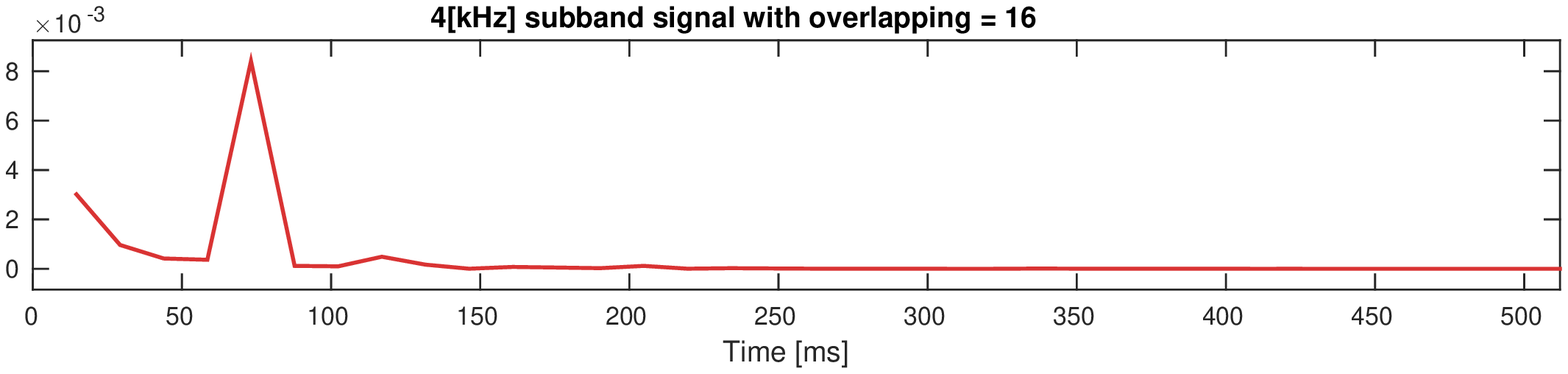}

\includegraphics[width=\textwidth]{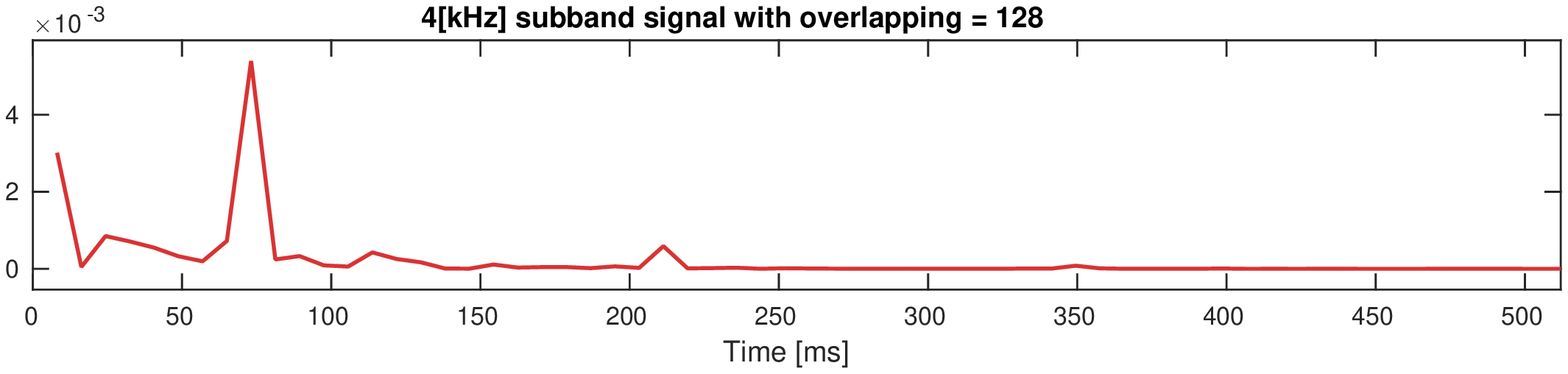}

\includegraphics[width=\textwidth]{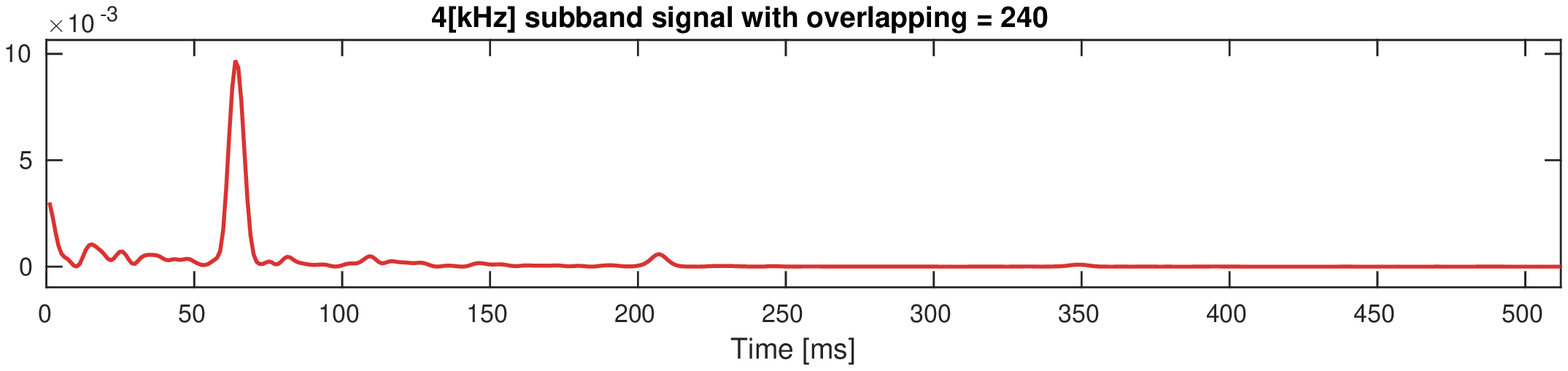}

\caption{Subband signals $H_{65}[n], \; n = 1,\ldots, N$, with  window length 256 and different overlappings. The signals show certain regularity, that increases with the window overlapping.}
\label{fi:H65s}
\end{figure}

In regards to $H$, although no general conditions are expected on its individual components, we do expect its first order time differences to exhibit a certain degree of regularity (see Figures \ref{fi:rirspecs} and \ref{fi:H65s}). In fact, if windows are set close enough relative to the duration of the reverberation phenomenon, then consecutive time components of $H$ will capture overlapped information, which along with the exponential decay characteristic of the RIR (\cite{ratnam2003blind}) accounts for a somewhat smooth structure. Therefore, we define the time differences matrix $V\in\mathbb{R}^{K\times(N_h-1)}$, with components $V_k[n] \doteq H_k[n]-H_k\left[n-1\right]\;\forall n=1,\ldots, N_h-1,\;k=1,\ldots, K$. The regularity of these variations is contemplated by  assuming $V$ follows a normal distribution:
\begin{equation}
\pi_{prior}(V) =  \prod_{k=1}^K \prod_{n=2}^{N_h} \frac{1}{\sqrt{2\pi}\eta_k }\exp \left(-\frac{V_k[n]^2}{\eta_k^2}\right). \nonumber
\end{equation}

Using Bayes' theorem, the \emph{a posteriori} joint distribution of $S$ and $H$ conditioned to $Y$ satisfies
\begin{equation} \label{eq:pipost}
\pi_{post}(S,H|Y)  \propto \pi_{like}(Y|S,H)\pi_{prior}(S)\pi_{prior}(H).
\end{equation}

%---------------------------------------------------------------------------------------------
\subsection{Mixed penalization}
%---------------------------------------------------------------------------------------------

Our goal is to find $\hat{S}$ and $\hat{H}$ that are representative of the \emph{a posteriori} distribution (\ref{eq:pipost}). Although the immediate instinct might be to compute the expected value, there are quite a few other ways to proceed, with different degrees of reliability and complexity. In lights of the assumed distributions and the high dimensionality of the problem, the \emph{maximum a posteriori} (MAP) estimator is a reasonable choice in this case. Note that maximizing (\ref{eq:pipost}) is tantamount to minimizing $-\log \pi_{post}(S,H|Y)$. If we denote by  $S_k,Y_k, X_k \in\mathbb{R}^N $, $H_k\in \mathbb{R}^{N_h}$ and $V_k\in \mathbb{R}^{N_h-1}$ the (transposed) rows of $S,Y, X, H$ and $V$, and define $L\in\mathbb{R}^{N_h-1\times N_h}$ in such a way that $LH_k = V_k$, then
\small
\begin{align} \nonumber
-\log \pi_{post}(S,H|Y)
& = \sum_{k=1}^{K}\left[  \frac{1}{\sigma^2}||Y_k - X_k||_2^2  +\frac{1}{b_k^p}\sum_n|S_k[n]|^p + \frac{1}{\eta_k^2}||LH_k||_2^2 \right] +C\nonumber,
\end{align}
\normalsize
where $C$ is a constant which does not depend on $S$ nor $H$.

Finally, the latter equation leads to the cost function
\begin{equation} \label{eq:cost-fun}
J(S,H) \doteq \sum_{k=1}^K\left( ||Y_k - X_k||_2^2  +\lambda_{s,k} ||S_k||_p^p + \lambda_{h,k}||LH_k||_2^2 \right),
\end{equation}
which shall be minimized to find our regularized solution. In this context,  $\lambda_{s,k}, \lambda_{h,k}\geq 0$ can be thought of as penalization parameters weighting both penalizers relative to the fidelity term, whereas the exponent $p\in (0,2)$ is a tunning parameter. It is timely to point out that small values of $p$ will promote sparsity, whereas values close to $2$ will promote smoothness. Since there is a clear scale indeterminacy in  the representation (\ref{eq:mod-rep}), we impose the (somewhat arbitrary) additional constraint $||S_k||_\infty=||Y_k||_\infty\;\forall k$, which means that the maximum values shall remain equal for every frequency.

%%-------------------------------
\subsection{Regularization parameters}
%%-------------------------------

As mentioned before, the parameters $\lambda_{h,k}, \lambda_{s,k},\; k=1,\ldots, K,$ weight the penalizers against the fidelity term. In this sense, the optimal weights of these regularization parameters might vary as a function of the frequency subband, and hence their proposed dependency on $k$. Since searching blindly for $2K$ parameters is non-viable in practice, we quantify this dependency by defining $\lambda_{h,k} \doteq \lambda_{h}\sum_{n=1}^N|Y_k[n]|^2$ and $\lambda_{s,k} \doteq \lambda_{s} \; \forall k = 1,\ldots,K$ (note that the relation between $S_k$ and $Y_k$ is already contemplated in the constraint that intends to avoid scale indeterminacy). This means we only need to look for two parameters ($\lambda_h, \lambda_s$) and then multiply $\lambda_{h}$ by the energy of the signal associated to each row of $Y$.

Next, we present an algorithm for approximating matrices $H$ and $S$ minimizing $J$.

%===============================================
\section{Updating rules}
%===============================================

We shall build an iterative algorithm following the idea in \cite{kameoka2009}, which is based on the auxiliary function technique.

Let $\Omega\subset\mathbb{R}$ and $f:\Omega\rightarrow\mathbb{R}_0^+$. Then, $g:\Omega\times\Omega\rightarrow\mathbb{R}_0^+$ is called an \emph{auxiliary function} for $f$ if
\begin{equation} \label{eq:aux_cond}
(i)\; g(w,w) = f(w)  \;\;\text{and}\;\;  (ii)\; g(w,w')\geq f(w), \;\; \forall w,w'\in\Omega.
\end{equation}

Let $w^0\in\Omega$ be arbitrary, and let
\begin{equation} \label{eq:gen-up-rule}
w^j \doteq \argmin_wg(w,w^{j-1}).
\end{equation}
With this definition, it can be shown (\cite{lee2001}) that the sequence $\{f(w^j)\}_j$ is non-increasing. We intend to use this property as a tool for alternatively updating the matrices $H$ and $S$. Let us begin by fixing $H=H'$, where $H'$ is an arbitrary $K\times N_h$ matrix. We will show that
\begin{align}\label{eq:aux_s}
g_s(S,S') \doteq &\sum_{k,n,\tau} \frac{S'_k[\tau]H'_k[n-\tau]}{X'_k[n]}\left( Y_k[n]-\frac{S_k[\tau]}{S'_k[\tau]}X'_k[n] \right)^2 +\sum_k\lambda_{h,k}||LH'_k||_2^2\nonumber \\ &+\sum_{k,n}\lambda_{s,k}\left(\frac{p}{2}S'_k[n]^{p-2}S_k[n]^2+|S'_k[n]|^{p} -\frac{p}{2}|S'_k[n]|^{p}\right)
\end{align}
is an auxiliary function for $J$ (as defined  in (\ref{eq:cost-fun})) with respect to $S$. From this point on, we denote by $X'_k[n] =  \sum_{\tau} S'_k[n-\tau] H'_k[\tau]$. The equality condition $(i)$ in (\ref{eq:aux_cond}) is rather straightforward. In fact,
\small
\begin{align}
g_s(S,S)
= &\sum_{k,n,\tau} \frac{S_k[\tau]H'_k[n-\tau]}{\sum_{\nu} S_k[\nu]H'_k[n-\nu]}\left( Y_k[n]-\frac{S_k[\tau]}{S_k[\tau]}\sum_{\nu} S_k[\nu]H'_k[n-\nu] \right)^2\nonumber \\
&+\sum_k\lambda_{h,k}||LH'_k||_2^2 +\sum_{k,n}\lambda_{s,k}\left(\frac{p}{2}S_k[n]^{p-2}S_k[n]^2+|S_k[n]|^{p} -\frac{p}{2}|S_k[n]|^{p}\right) \nonumber\\
=&\sum_{k,n,\tau} \frac{S_k[\tau]H'_k[n-\tau]}{\sum_{\nu} S_k[\nu]H'_k[n-\nu]}\left( Y_k[n]-\sum_{\nu} S_k[\nu]H'_k[n-\nu] \right)^2 \nonumber \\ &+\sum_k\lambda_{h,k}||LH'_k||_2^2  +\sum_{k,n}\lambda_{s,k}|S_k[n]|^{p} \nonumber \\
=&\sum_{k,n} \left( Y_k[n]-\sum_{\nu} S_k[\nu]H'_k[n-\nu] \right)^2 +\sum_k\lambda_{h,k}||LH'_k||_2^2+\sum_{k,n}\lambda_{s,k}|S_k[n]|^{p} \nonumber \\ =& J(S,H').\nonumber
\end{align}
\normalsize

To prove condition $(ii)$ in (\ref{eq:aux_cond}) we begin by defining
\begin{align}\nonumber
P_{k,n} \doteq& \sum_{\tau} \frac{S'_k[\tau]H'_k[n-\tau]}{X'_k[n]}\left( Y_k[n]-\frac{S_k[\tau]}{S'_k[\tau]}X'_k[n] \right)^2, \nonumber\\
R_{k,n} \doteq& ( Y_k[n]-\sum_{\tau}S_k[\tau]H'_k[n-\tau] )^2, \nonumber
\end{align}
 and $Q:\mathbb{R}^+\rightarrow \mathbb{R}$ such that $Q(x) \doteq \frac{p}{2}x^{p-2}S_k[n]^2+x^p -\frac{p}{2}x^{p}$. With these definitions, we can write

\begin{equation} \nonumber
g_s(S,S') = \sum_{k}\left(\sum_{n}(P_{k,n} +\lambda_{s,k}Q(S'_k[n]) )+\lambda_{h,k}||LH'_k||_2^2\right),
\end{equation}
and
\begin{equation} \nonumber
J(S,H') =  \sum_{k}\left(\sum_{n}(R_{k,n} +\lambda_{s,k}|S_k[n]|^p)+\lambda_{h,k}||LH'_k||_2^2\right).
\end{equation}

Hence, to prove that $g_s(S,S') \geq J(S,H') \;\forall S, S'$ it is sufficient to show that $P_{k,n} \geq R_{k,n}$ and $Q(S'_k[n])\geq |S_k[n]|^p \; \forall n = 1,\ldots, N, k=1,\ldots, K$. In fact,
\small
\begin{align}
P_{k,n}-R_{k,n}
=& \sum_{\tau} \frac{S'_k[\tau]H'_k[n-\tau]}{X'_k[n]}\left( Y_k[n]-\frac{S_k[\tau]}{S'_k[\tau]}X'_k[n] \right)^2 \nonumber\\&-( Y_k[n]-\sum_{\tau}S_k[\tau]H'_k[n-\tau] )^2 \nonumber\\
=& \sum_{\tau}\frac{H'_k[n-\tau]S_k[\tau]^2X'_k[n]}{S'_k[\tau]} - \left(\sum_{\tau}S_k[\tau]H'_k[n-\tau]\right)^2\nonumber\\
=& \sum_{\tau,\nu}\frac{H'_k[n-\tau]S_k[\tau]^2H'_k[n-\nu]S'_k[\nu]}{S'_k[\tau]} - \sum_{\tau,\nu}S_k[\tau]H'_k[n-\tau]S_k[\nu]H'_k[n-\nu] \nonumber\\
=&\sum_{\tau,\nu}\left(  \frac{H'_k[n-\tau]S_k[\tau]^2H'_k[n-\nu]S'_k[\nu]}{S'_k[\tau]} - S_k[\tau]H'_k[n-\tau]S_k[\nu]H'_k[n-\nu]\right) \nonumber\\
=& \sum_{\tau\neq \nu}\left(  \frac{H'_k[n-\tau]S_k[\tau]^2H'_k[n-\nu]S'_k[\nu]}{S'_k[\tau]} - S_k[\tau]H'_k[n-\tau]S_k[\nu]H'_k[n-\nu]\right) \nonumber\\
=& \sum_{\tau<\nu}H'_k[n-\tau]H'_k[n-\nu]\left(  \frac{S_k[\tau]^2S'_k[\nu]}{S'_k[\tau]} - 2S_k[\tau]S_k[\nu] +\frac{S_k[\nu]^2S'_k[\tau]}{S'_k[\nu]}\right) \nonumber\\
=& \sum_{\tau<\nu}\frac{H'_k[n-\tau]H'_k[n-\nu]}{S'_k[\nu]S'_k[\tau]}\left(S_k[\tau]S_k'[\nu]-S_k[\nu]S_k'[\tau]\right)^2 \geq 0. \label{eq:A_B} \nonumber
\end{align}
\normalsize

To prove that  $Q(S'_k[n])\geq |S_k[n]|^p$, we begin by noting that $Q\in\mathcal{C}^{\infty}(\mathbb{R}^+)$. Then, the first order necessary condition for $Q$ yields
\begin{equation}\nonumber
0 = \frac{\partial Q}{\partial x} = \frac{p(p-2)}{2}x^{p-3}S_k[n]^2+px^{p-1} -\frac{p^2}{2}x^{p-1}= \frac{p(p-2)}{2}x^{p-1}(x^{-2} S_k[n]^2-1),
\end{equation}
meaning the only point at which the derivative of $Q$ equals zero is at $x=S_k[n]$.   Furthermore, $\frac{\partial^2}{\partial x^2}Q( S_k[n])= S_k[n]^{p-2} (2p-p^2)>0 \;\forall p\in(0,2)$, meaning that  $Q(S_k[n]) = |S_k[n]|^p$ is the global minimum of $Q$. This yields
\begin{align} \nonumber
g_s(S,S') & = \sum_{k}\left(\sum_{n}(P_{k,n} +\lambda_{s,k}Q(S'_k[n]) )+\lambda_{h,k}||LH'_k||_2^2\right) \\ \nonumber
& \geq  \sum_{k}\left(\sum_{n}(R_{k,n} +\lambda_{s,k}|S_k[n]|^p)+\lambda_{h,k}||LH'_k||_2^2\right) = J(S,H').
\end{align}
\begin{flushright}
$\blacksquare$
\end{flushright}

In an analogous way, it can be shown that if we let $S=S'$ be fixed, where $S'$ is an arbitrary $K\times N$ matrix, then
\begin{align}\nonumber %\label{eq:aux_h}
g_h(H,H') \doteq &\sum_{k,n,\tau} \frac{S'_k[n-\tau]H'_k[\tau]}{X'_k[n]}\left( Y_k[n]-\frac{H_k[\tau]}{H'_k[\tau]}X'_k[n] \right)^2 \nonumber\\
&+\sum_{k}\lambda_{s,k} ||S'_k||_p^p +\sum_{k} \lambda_{h,k}||LH_k||_2^2 \nonumber
\end{align}
is an auxiliary function for $J(S',H)$ with respect to $H$.

Having defined auxiliary functions, we will use the updating rule derived from (\ref{eq:gen-up-rule}) to build an algorithm for iteratively approaching matrices $S$ and $H$ minimizing $J$. Notice this requires minimizing $g_s$ and  $g_h$ with respect to the updating variables, but since $g_s$ is quadratic with respect to $S$  and $g_h$ is quadratic with respect to $H$, we can simply use the first order necessary conditions in both cases. From this point on, in the context of the iterative updating process,  $S'$ and $H'$ will refer not to arbitrary nonnegative matrices, but to those estimations of $S$ and $H$ obtained in the immediately previous step.

%%-------------------------------
\subsection{Updating rule for S}
%%-------------------------------

Firstly, we shall derive an updating rule for $S_k[\tau]$. That is, we wish to minimize $g_s$ w.r.t. $S$. The first order necessary condition on $g_s$ yields

\small
\begin{align}
0 = & \frac{\partial g_s(S,S')}{\partial S_k[\tau] } \nonumber \\
 = & -2 \sum_n H'_k[n-\tau]\left(Y_k[n]-\frac{S_k[\tau]}{S_k'[\tau]}X'_k[n]\right)+\lambda_{s,k}pS_k'[\tau]^{p-2}S_k[\tau] \nonumber \\
 = & -\sum_n H'_k[n-\tau]Y_k[n]+\frac{S_k[\tau]}{S_k'[\tau]}\sum_n H'_k[n-\tau] X'_k[n]+\frac{\lambda_{s,k}}{2}pS_k'[\tau]^{p-2}S_k[\tau] \nonumber \\
 = & -S_k'[\tau] \sum_n H'_k[n-\tau]Y_k[n]+\left(\sum_n H'_k[n-\tau] X'_k[n] +\frac{\lambda_{s,k}}{2}pS_k'[\tau]^{p-1}\right)S_k[\tau], \nonumber
\end{align}
\normalsize
which easily leads to the multiplicative updating rule
\begin{equation} \nonumber
S_k[\tau] = S_k'[\tau]\frac{\sum_n H'_k[n-\tau]Y_k[n]}{\sum_n H'_k[n-\tau]X'_k[n] \,+\,\frac{\lambda_{s,k}}{2}p|S'_k[\tau]|^{p-1}}.
\end{equation}
In order to avoid the aforementioned scale indeterminacy, every updating step is to be followed by scaling $S_k$ so that its $\ell^\infty$ norm coincides with that of the observation $Y_k$.

%%-------------------------------
\subsection{Updating rule for H}
%%-------------------------------

In order to find an updating rule for $H$, we shall write $g_h$ as a function of the transposed rows $H_k$. We begin by noting
\begin{align}\nonumber
g_h(H,H') = &\sum_{k,n,\tau} \frac{S'_k[n-\tau]H'_k[\tau]}{X'_k[n]}\left( Y_k[n]-\frac{H_k[\tau]}{H'_k[\tau]}X'_k[n] \right)^2 \nonumber \\
&+\sum_{k}\lambda_{s,k} ||S'_k||_p^p +\sum_{k} \lambda_{h,k}||LH_k||_2^2 \nonumber \\
= &\sum_{k,n,\tau}\frac{S'_k[n-\tau]H'_k[\tau]Y_k^2[n]}{X'_k[n]} -2\sum_{k,n,\tau}S'_k[n-\tau]Y_k[n]H_k[\tau] \nonumber \\  &+\sum_{k,n,\tau}\frac{S'_k[n-\tau]X'_k[n]H^2_k[\tau]}{H'_k[\tau]}\nonumber \\
&+\sum_{k}\lambda_{s,k} ||S'_k||_p^p +\sum_{k} \lambda_{h,k}||LH_k||_2^2. \nonumber
\end{align}

Next, we define the diagonal matrices $A^k,B^k\in\mathbb{R}^{N_h\times N_h}$, whose diagonal elements are $A^k_{\tau,\tau} \doteq \sum_{n}S'_k[n-\tau]X'_k[n]$ and $B^k_{\tau,\tau} \doteq H'_k[\tau]$, and the vector  $\zeta^k\in\mathbb{R}^{N_h}$ with components $\zeta^k_\tau=\sum_{n}S'_k[n-\tau]Y_k[n]$. With these definitions, we can write
\begin{align}\nonumber
g_h(H,H') = &\sum_{k,n,\tau} \frac{S'_k[n-\tau]H'_k[\tau]Y_k^2[t]}{X'_k[n]} -2\sum_k H_k\T  \zeta^k +\sum_k H_k\T  A^k(B^k)^{-1}H_k\\ \nonumber
&+\sum_{k}\lambda_{s,k} ||S'_k||_p^p +\sum_{k} \lambda_{h,k}H_k\T  L\T  LH_k.
\end{align}

Now, the first order necessary condition for $g_h$ with respect to $H_k$ is given by
\begin{align}
0 = \frac{\partial g_h(H,H')}{\partial H_k}  = -2 \zeta^k +2A^k(B^k)^{-1}H_k  +2 \lambda_{h,k}L\T  LH_k,
\end{align}
which readily leads to an updating rule consisting of solving the linear system
\begin{equation} \label{eq:sistH}
(A^k+\lambda_{h,k}B^kL\T L)H_k = B^k\zeta^k.
\end{equation}

Let us notice that under the assumption that the diagonal elements of $A^k$ and $B^k$ are strictly positive, and since $L\T L$ is positive-semidefinite, $(B^k)^{-1}A^k +\lambda_{h,k}L\T L$ is positive-definite, and hence the linear system has a unique solution. The assumption of $A^k_{\tau,\tau}>0$ is adequate, since these elements correspond to the discrete convolution of $S'_k$ and $X'_k$. Although the validity of the hypothesis over $B^k_{\tau,\tau}$ is not so clear, in practice, the matrix in system (\ref{eq:sistH}) has turned out to be non-singular. Nonetheless, $H_k$ can be computed as the best approximate solution in the least-squares sense. Then, solving this $N_h\times  N_h$  linear system entails no challenge, since $N_h$ is usually chosen relatively small, depending on the window step and the reverberation time.

All the steps for the dereverberation process are stated in Algorithm \ref{al:mixpen}. Note that in the initialization we define the clean spectrogram $S$ equal to the observation, which is natural since in a way they both correspond to the same signal, and $H_k$ as a vector with exponential time decay, which is an expected characteristic of a RIR. Finally, we set the stopping criterion over the decay of the norm of two consecutive approximations of $S$. This has shown to work quite well, although other stopping criteria might be considered.

Results to illustrate the performance of the algorithm are presented in the next section.

\begin{algorithm}
\caption{Mixed penalization dereverberation}
\label{al:mixpen}
\begin{algorithmic}[1]
\NoDo
\NoThen
\STATE  \textbf{Initializing}
\STATE $S \leftarrow Y$
\STATE $H_k[n] \leftarrow \exp( -n)$  \;\;  $ \forall k = 1\ldots K, \;n = 1\ldots N$
\STATE $ $

\STATE \textbf{MAIN LOOP}

\FOR{$i=1\ldots \text{maxiter}$}
	\vspace{0.4cm}
	\STATE $S' = S.$
	\vspace{0.4cm}

	\STATE $\displaystyle{X_k[n] \leftarrow \sum_{\tau} S_k[n-\tau] H_k[\tau]} \;\; \;\; \forall k = 1\ldots K, \;n = 1\ldots N$
	\vspace{0.4cm}	

	\FOR{$k = 1\ldots K$}
		\FOR{$\tau = 1\ldots N$}
			\STATE $\displaystyle{S_k[\tau] \leftarrow S_k[\tau]\frac{\sum_n H_k[n-\tau]Y_k[n]}{\sum_n H_k[n-\tau]X_k[n] \,+\,\frac{\lambda_{s,k}}{2}p|S_k[\tau]|^{p-1}}}.$
		\ENDFOR
		\STATE $\displaystyle{S_k \leftarrow S_k \frac{\|Y_k\|_{\infty}}{\|S_k\|_{\infty}}.}$
	\ENDFOR
	\vspace{0.4cm}
	
	\FOR{$k = 1\ldots K$}
		\STATE$\text{Build the diagonal matrices} A^k,B^k\in\mathbb{R}^{N_h\times N_h}:$
		\STATE$\;\;\; A^k_{\tau ,\tau} = \sum_{n}S_k[n-\tau]X_k[n],$
		\STATE$\;\;\; B^k _{\tau ,\tau} = H_k[\tau].$
		\STATE$\text{Build the vector } \zeta^k:$
		\STATE$\;\;\; \zeta^k_\tau = \sum_{n}S_k[n-\tau]Y_k[n]$
		\STATE$\text{Solve for } H_k:$
		\STATE$\;\;\;  \displaystyle{(A^k+\lambda_{h,k}B^kL\T L)H_k = B^k\zeta^k.}$
	\ENDFOR
	\vspace{0.4cm}

	\IF{$\|S-S'\|_F\leq \delta$}
		\RETURN
	\ENDIF
	\vspace{0.4cm}
\ENDFOR

\end{algorithmic}
\end{algorithm}

%===============================================
\section{Experimental results}
%===============================================

For the experiments, we took $110$ speech signals from the TIMIT database (\cite{zue1990timit}), recorded at 16 KHz,  and artificially made them reverberant by convolution with impulse responses generated with the software Room Impulse Response Generator\footnote{https://github.com/ehabets/RIR-Generator}, based on the model in \cite{allen1979}. Each signal was degraded under different reverberation conditions: three different room sizes, each with three different microphone positions and four different reverberation times.\footnote{A web demo for our algorithm can be found in http://fich.unl.edu.ar/sinc/web-demo/blindder/}

In order to avoid preprocessing, the choice of the regularization parameters was made \emph{a priori} by means of empirical rules, based upon signals from a different database. This is supported by the fact that the parameters were observed to be rather robust with respect to variations of the reverberation conditions, and hence they were chosen simply as $\lambda_h = 1$ and $\lambda_s = 10^{-4}$. The rest of the model parameters were chosen as specified in Table \ref{tab:params}.

\begin{table}[h]
\centering
\begin{tabular}{|ccccccc|}
\hline
 $p$ & $N_h$ & win. & window size & win. overlap. & $\delta$ &max. iter. \\
\hline
\hline
1 & 15 & Hann & 512 samples& 256 samples & \small{$\|Y\|_F\times 10^{-3}$} &20 \\
\hline
\end{tabular}
\caption{Model parameter values}
\label{tab:params}
\end{table}

Let us point out that the choice of $N_h$ was done as to allow $H$ to capture early reverberation while precluding overlapped representations. In the first place, it is desirable for $H$ to represent the RIR along the full Early Decay Time (EDT), the time period in which the reverberation phenomenon alters the clean signal the most, so its effect can be effectively nullified. On the other hand, if we were to choose $N_h$ too large, it might lead certain similarities in the observation $Y$ within a fixed frequency range to be represented as echoes from high energy components of $S$. It is worth mentioning, however, that the performance of our dereverberation method has shown no high sensitivity with respect to the choice of $N_h$.

In order to evaluate the performance of our model, we made comparisons against two state of the art methods that work under the same conditions. The one proposed by Kameoka \emph{et al} in \cite{kameoka2009}, choosing all the parameters as suggested, and the one proposed by Wisdom \emph{et al} in \cite{wisdom2014}, with a window length of $2048$.

\begin{figure}[h]
\centering
\includegraphics[width=0.8\textwidth]{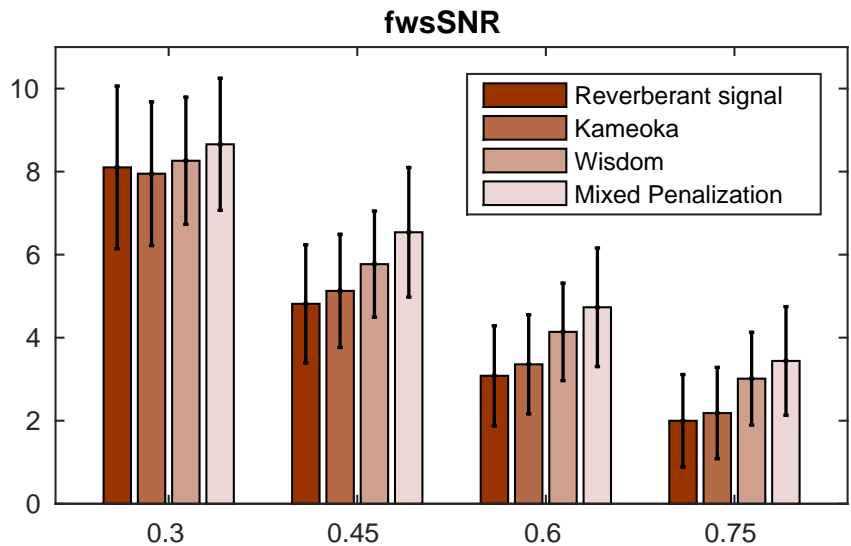}
\caption{Mean and standard deviations of performance fwsSNR for different reverberation times.}
\label{fi:fwsSNR}
\end{figure}

\begin{figure}[h]
\centering
\includegraphics[width=0.8\textwidth]{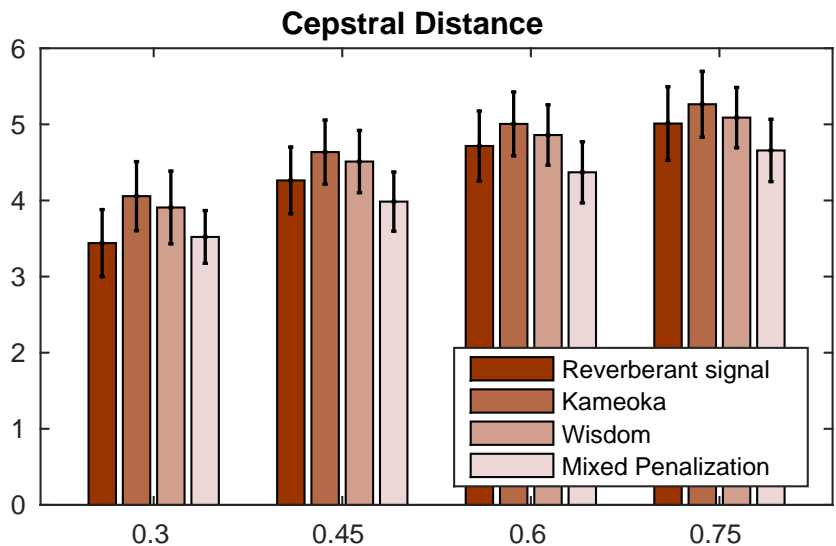}
\caption{Mean and standard deviations of cepstral distance for different reverberation times.}
\label{fi:CD}
\end{figure}

\begin{figure}[h]
\centering
\includegraphics[width=0.8\textwidth]{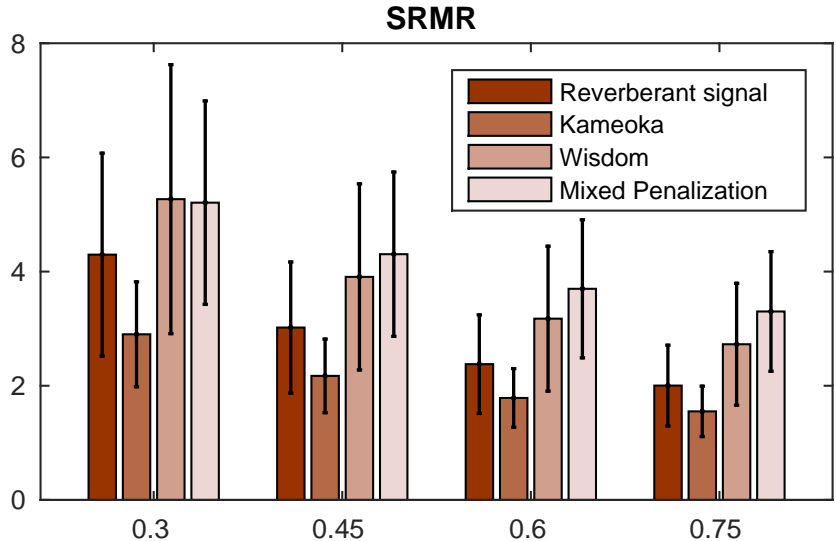}
\caption{Mean and standard deviations of SRMR for different reverberation times.}
\label{fi:SRMR}
\end{figure}

To measure performance, following \cite{hu2008}, we made use of the frequency weighted segmental  signal-to-noise ratio (fwsSNR) and cepstral distance. Furthermore, we also measured the  speech-to-reverberation modulation energy ratio (SRMR, \cite{falk2010}), which has the advantage of being non-intrusive (it does not use the clean signal as an input). The results for each performance measure are stated in Tables \ref{tab:results_SNR}-\ref{tab:results_SRMR}  and  depicted in Figures \ref{fi:fwsSNR}- \ref{fi:SRMR}, classified in function of the reverberation times: $300[\text{ms}]$, $450[\text{ms}]$, $600[\text{ms}]$ and $750[\text{ms}]$. Notice that for the cases of fwsSNR and SRMR, higher values correspond to better performance, while for the cepstral distance, small values indicate higher quality.

\begin{table}[h]
\small
\centering
\begin{tabular}{| C{1.9cm} | C{2.4cm}  | C{2.4cm} C{2.4cm} C{2.4cm} |}
\hline
Rev. time [ms] & Rev. Signal & Kameoka der. & Wisdom der. & Mixed penalization \\
\hline
\hline
300 & 8.102 (1.96)& 7.950 (1.73)& 8.262 (1.53)& \textbf{8.658} (1.59)\\
450 & 4.815 (1.42)& 5.127 (1.36)& 5.771 (1.28)& \textbf{6.539} (1.56)\\
600 & 3.082 (1.20)& 3.358 (1.19)& 4.140 (1.17)& \textbf{4.732 }(1.43)\\
750 & 1.998 (1.11)& 2.184 (1.10)& 3.013 (1.12)& \textbf{3.440} (1.31)\\
\hline
\end{tabular}\normalsize
\caption{Mean and (standard deviation) of fwsSNR for each method and reverberation time (best results in boldface).}
\label{tab:results_SNR}
\end{table}

\begin{table}[h]
\small
\centering
\begin{tabular}{| C{1.9cm} | C{2.4cm}  | C{2.4cm} C{2.4cm} C{2.4cm} |}
\hline
Rev. time [ms] & Rev. Signal & Kameoka der. & Wisdom der. & Mixed penalization \\
\hline
\hline
300 & 3.440 (0.44)& 4.057 (0.45)& 3.908 (0.48)&\textbf{ 3.521} (0.35)\\
450 & 4.264 (0.44)& 4.636 (0.42)& 4.511 (0.41)& \textbf{3.985 }(0.39)\\
600 & 4.716 (0.46)& 5.006 (0.42)& 4.860 (0.40)& \textbf{4.370} (0.40)\\
750 & 5.011 (0.48)& 5.264 (0.43)& 5.089 (0.40)& \textbf{4.657} (0.41)\\
\hline
\end{tabular}\normalsize
\caption{Mean and (standard deviation) of cepstral distance for each method and reverberation time (best results in boldface).}
\label{tab:results_CD}
\end{table}

\begin{table}[h]
\small
\centering
\begin{tabular}{| C{1.9cm} | C{2.4cm}  | C{2.4cm} C{2.4cm} C{2.4cm} |}
\hline
Rev. time [ms] & Rev. Signal & Kameoka der. & Wisdom der. & Mixed penalization \\
\hline
\hline
300 & 4.297 (1.78)& 2.901 (0.92)& \textbf{5.269} (2.36)& 5.207 (1.78)\\
450 & 3.020 (1.15)& 2.173 (0.64)& 3.907 (1.63)& \textbf{4.305 }(1.44)\\
600 & 2.378 (0.86)& 1.786 (0.51)& 3.175 (1.27)& \textbf{3.698 }(1.21)\\
750 & 2.003 (0.71)& 1.551 (0.44)& 2.727 (1.07)& \textbf{3.301} (1.05)\\
\hline
\end{tabular}\normalsize
\caption{Mean and (standard deviation) of SRMR for each method and reverberation time (best results in boldface).}
\label{tab:results_SRMR}
\end{table}

In regard to the fwsSNR performance measure, the values in Table \ref{tab:results_SNR} (Figure \ref{fi:fwsSNR}) give account of significant improvements of our proposed method with respect to the other two. This improvement becomes more evident as the reverberation time increases. As for the cepstral distance, although the results in Table \ref{tab:results_CD} (Figure \ref{fi:CD}) account for a better performance of our proposed method, the quality with respect to the reverberant signal is improved only for reverberation times of 450[ms] or more. Finally, the SRMR also shows an improvement with respect to the other methods for reverberation times of  450[ms] or greater (see Table \ref{tab:results_SRMR}, Figure \ref{fi:SRMR}).

%===============================================
\section{Conclusions}
%===============================================

In this work, a new blind dereverberation method for speech signals based on regularization over a convolutive NMF representation of the signal spectrograms was introduced and tested. Results show a significant improvement over the state of the art methods, specially for high reverberation times. There is certainly much room for improvement, e.g. finding ways of optimally choosing the regularization parameters, exploring the use of other penalizers, etc.

%===============================================
\section*{Acknowledgements}
%===============================================

This work was supported in part by Consejo Nacional de Investigaciones Cient\'ificas y T\'ecnicas, CONICET  through PIP 2014-2016 N$^\text{o}$ 11220130100216-CO, \hspace{0.1cm} the Air Force Office of Scientific Research,  AFOSR/SOARD, through Grant FA9550-14-1-0130, by Universidad Nacional del Litoral, UNL, through CAID-UNL 2011 N$^\text{o}$ 50120110100519 ``Procesamiento de Se\~nales Biom\'edicas.'' and CAI+D-UNL 2016, PIC 50420150100036LI ``Problemas Inversos y Aplicaciones a Procesamiento de Se\~nales e Im\'agenes''.




 \bibliographystyle{elsarticle-num}
 \bibliography{ref_der}

\end{document}